\documentclass[prb,twocolumn,reprint,showpacs,superscriptaddress,floatfix]{revtex4-1}
\usepackage{epsfig}
\begin{document}

\title{Antiferromagnetic Exchange Interactions in Ni$_{2}$Mn$_{1.4}$In$_{0.6}$ ferromagnetic Heusler alloy}

\author{K. R. Priolkar}
\email[corresponding author: ]{krp@unigoa.ac.in}
\author{P. A. Bhobe}\altaffiliation{School of Basic Sciences, Indian Institute of Technology Indore, Indore, 452017 India}
\author{D. N. Lobo}
\affiliation{Department of Physics, Goa University, Taleigao Plateau, Goa 403206, India}
\author{S. W. D'Souza}
\author{S. R. Barman}
\affiliation{UGC-DAE Consortium for Scientific Research, Khandwa Road, Indore, 452001, Madhya Pradesh, India}
\author{Aparna Chakrabarti}
\affiliation{Raja Ramanna Centre for Advanced Technology, Indore 452013, Madhya Pradesh, India}
\author{S. Emura}
\affiliation{Institute of Scientific and Industrial Research, Osaka University, 8-1 Mihogaoka, Ibaraki, Osaka 567-0047, Japan}

\date{\today}

\begin{abstract}
Magnetism in Ni-Mn-Z (Z = Ga,In,Sn,Sb) Heusler alloys has so far been predominantly attributed to Rudermann-Kittel-Kasuya-Yoshida type  interactions between Mn atoms. We investigate magnetic interactions in one such  alloy, Ni$_{2}$Mn$_{1.4}$In$_{0.6}$ and attempt to explain the origin of antiferromagnetic (AFM) interactions that coexist with ferromagnetic ones. Through the combination of x-ray absorption spectroscopy and x-ray magnetic circular dichroism (XMCD), we find that Ni plays an important role along with Mn in the overall magnetism.  A significant hybridization that develops between Mn and Ni orbitals results in a small antiferromagnetic moment at Ni sites. The shift in the XMCD hysteresis loops in the martensitic phase suggests that antiferromagnetism results from superexchange like interactions between Mn atoms mediated by Ni.

\end{abstract}
\maketitle

\section{Introduction}
Mn-rich Heusler alloys of the type Ni$_2$Mn$_{1+x}$Z$_{1-x}$ (Z=In, Sn, Sb) exhibit interesting properties like inverse magnetocaloric effect, large magnetic field induced strain, giant magnetoresistance and exchange bias effect \cite{planes, acet, kai, li, pat}. The origin of these effects lies in the coupling between martensitic structural transition and magnetic degrees of freedom of these alloys. The high temperature ($T$) austenitic phase is ferromagnetic (FM), which arises due to Rudermann-Kittel-Kasuya-Yoshida (RKKY) exchange interactions between Mn atoms. However, the magnetism of the martensitic phase is till date elusive. Polarized neutron scattering experiments describe this phase as antiferromagnetic (AFM) \cite{aksoy}, whereas M\"{o}ssbauer study indicates it to be paramagnetic (PM) in nature \cite{khovailo}. Agreement, however, exists on the presence of a strong competition between FM and AFM interactions, but the origin of AFM interactions remains unclear. Recent observation of spin-valve like magnetoresistance in Mn$_2$NiGa \cite{barman-PRL}, ab initio calculations of magnetic exchange parameters of Ni$_2$Mn$_{1+x}$Sn$_{1-x}$ \cite{entel-PRB} and Monte Carlo simulations of Ni$_2$Mn$_{1+x}$Z$_{1-x}$ \cite{buchel}, indicate that structural disorder in the Mn site occupancy influences the magnetic properties of these compounds. However, these calculations do not take into account the local structural distortions which have been shown to be present in Mn-rich compositions of Ni-Mn-Z alloys \cite{krp-apl}.

Monte carlo simulations indicate the origin of AFM in Ni$_2$Mn$_{1+x}$Z$_{1-x}$ is due to interactions between Mn atoms at their own sub-lattice (Mn$_{\rm Mn}$) and those occupying Z sub-lattice (Mn$_{\rm Z}$) \cite{buchel}. Alternately, first principle calculations by E. \c{S}a\c{s}\i o\u{g}lu et al \cite{sasi} emphasize that AFM superexchange interactions become prominent when the unoccupied Mn 3d band lies closer to the Fermi level ($E_F$). In this regard, the Ni-Mn hybridization and local structural distortions gain relevance as these processes can affect Mn-band position in the overall electronic structure. Recent EXAFS study demonstrates a one-to-one correspondence between  temperature dependent change in Ni-Mn bond distance and magnetization of Ni$_2$Mn$_{1.4}$In$_{0.6}$, thus reinforcing such a view \cite{krp-epl}. Full potential linearized augmented plane wave (FPLAPW) calculations stress on the importance of Ni-Mn hybridization in stabilizing a ferrimagnetic ground state in Mn$_2$NiGa/In \cite{ac-epl}. Therefore, the present study aims at understanding the origin of AFM and the role played by each constituent atom in the magnetism of these Mn rich Heusler compositions. A combination of x-ray absorption spectroscopy (XAS) and x-ray magnetic circular dichroism (XMCD) measurements at the Mn and Ni L edges can serve as a perfect tool, as demonstrated by earlier studies on Ni$_2$MnZ (Z = Ga, In, Sn) alloys \cite{imada}. While XAS gives a picture of local unoccupied density of states, XMCD elucidates the local magnetism of the absorbing atom.

In the present study we make an attempt to understand the nature of magnetic interactions between Mn and Ni in the martensitic phase of Ni$_{2}$Mn$_{1.4}$In$_{0.6}$. We present  temperature dependent XAS and XMCD measurements of two samples: Ni$_{2}$MnIn and Ni$_{2}$Mn$_{1.4}$In$_{0.6}$ and supplement our results with {\it ab initio} spin polarized relativistic Korringa-Kohn-Rostoker (SPRKKR) Green's function calculations. Ni$_2$MnIn is a ferromagnet with Curie temperature, T$_C$ $\sim$ 306 K; it crystallizes in L2$_1$ crystal structure and does not undergo martensitic transformation. It is chosen here for its ferromagnetically ordered ground state with a stable crystal structure and prototypical Heusler composition. Substituting In by Mn to realize Ni$_{2}$Mn$_{1.4}$In$_{0.6}$ results in martensitic transformation in the region 250-295 K. A PM to FM transition at (T$_C$)$_A$=310 K in its austenitic phase is followed by another magnetic transition at (T$_C$)$_M$=200 K in its martensitic phase. We establish that strengthening of Ni-Mn hybridization in the region of martensitic transformation leads to \textit{Mn-Ni-Mn} type superexchange AFM interactions.

\section{Methods}
Polycrystalline samples used in the present study were prepared and characterized as described in Ref.\onlinecite{krp-epl}. The elemental compositions obtained from SEM-EDS were Ni = 50.1, Mn = 25.05, In = 24.85 for Ni$_2$MnIn and Ni = 50.25, Mn = 34.5, In = 15.25 for Ni$_2$Mn$_{1.4}$In$_{0.6}$. We performed polarization
dependent XAS measurements at BL25SU beamline at SPring8, Japan, using a total electron yield detection method \cite{naka}. The samples were fractured \textit{in situ} and a vacuum of $\sim 10^{-8}$ Torr was maintained throughout the experiment. X-rays were tuned to record the Mn and Ni L edges in the range, $T$= 15\,K to 310\,K. An external magnetic field up to 2T was applied in the direction parallel to the x-ray beam. The spectra were recorded for the positive and negative helicities of the circularly polarized x-rays. XAS signal was then extracted as the sum of positive ($\mu^+$) and negative ($\mu^-$) absorption coefficients, while XMCD was extracted as the difference between $\mu^+$ and $\mu^-$. After subtraction of a constant background in the pre-edge region, the XAS spectra were normalized with respect to the area under the curve. We also recorded the In M edge in both the samples but no XMCD signal was observed. The spin ($\mu_{spin}$) and orbital ($\mu_{orb}$) moments were extracted from XMCD data using the standard sum rules\cite{chen}.

In the SPRKKR calculation\cite{Ebert, Ebert11}, the number of $k$ points for SCF cycles were taken to be 500 in the irreducible BZ. The angular momentum expansion up to l$_{\rm max}$=3 has been used for each atom. The exchange and correlation effects were incorporated using the LDA framework \cite{Vosko80}. The L2$_1$ structure for Ni$_2$MnIn with Fm$\overline{3}$m space group and $a$=6.0537\AA~ is well known \cite{Ahuja10}. For Ni$_2$Mn$_{1.4}$In$_{0.6}$, the low temperature crystal structure is not fully established. Hence, we consider a simple tetragonal structure derived from the lattice parameters of the 10M modulated monoclinic cell as reported in literature \cite{Krenke06}, with lattice constants: $a_T$= \,[($a+c$/5)$\times$$\sqrt{2}$)]/2=\,6.1007\AA~ and $c_T$=\,$b$=\,5.882\AA.

\section{Results and Discussion}
\begin{figure}
\centering
\includegraphics[width=\columnwidth]{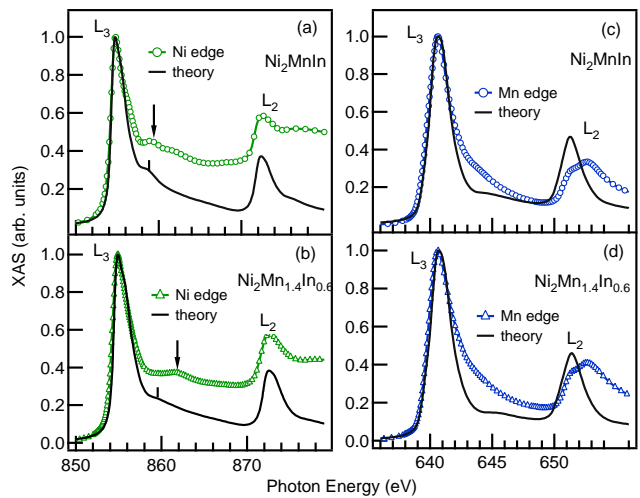}
\caption{(Color online) Ni and Mn $L_{2,3}$-edge XAS of the two compositions, measured at 15 K and compared with the calculated spectra. For better representation of the experimental and calculated spectra, the peak heights of L3 have been matched to unity.} \label{xas-Ni2MnIn}
\end{figure}

\begin{figure}[htb]
\centering \includegraphics[width=0.9\columnwidth]{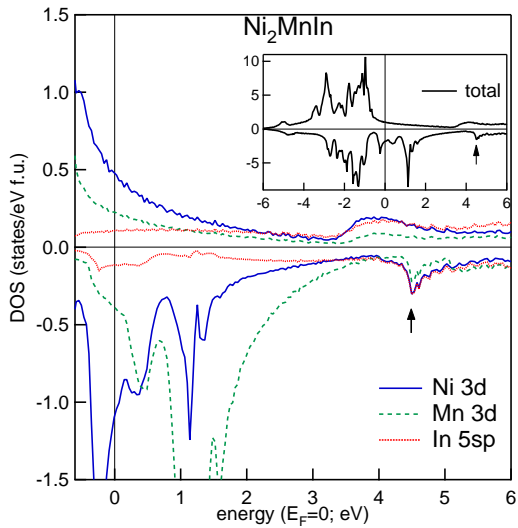} \caption{(Color online)  Angular momentum and spin projected unoccupied partial DOS of Ni$_2$MnIn. Inset shows the spin polarized total DOS over an extended range including occupied region below $E_F$. The DOS is in agreement with earlier studies
\cite{Ahuja10}, and the peak in the minority spin DOS at 1.1 eV arises from Mn 3d states.} \label{dosIn25}
\end{figure}

Figure \ref{xas-Ni2MnIn} presents XAS plots at $T$ = 15 K recorded at Mn and Ni L$_{2,3}$ edges in Ni$_2$MnIn and Ni$_2$Mn$_{1.4}$In$_{0.6}$ and, compared with the calculated spectra. A good agreement is obtained between experimental and the calculated spectra. The spectra recorded at various temperatures (15\,K to 310\,K) are presented in the supplementary text \cite{supplement}. Ni-XAS of Ni$_2$MnIn in Fig.~\ref{xas-Ni2MnIn}(a), exhibits a peak at 854.8\, eV and a shoulder at 856.5\, eV, that appears due to transition from Ni 2$p$ $\rightarrow$ 3$d$ states present above $E_F$. In addition, a satellite feature observed at 859.2\, eV (indicated with an arrow) that is, $~$4.4 eV above the L$_3$ edge. This feature is nicely reproduced in our calculated spectrum as well. Comparing the experimental spectrum with the minority spin DOS of Ni$_2$MnIn shown in Fig. \ref{dosIn25}, we find that the satellite feature corresponds with the peak at around 4.5 eV above E$_F$. This peak arises primarily from Ni 3$d$ - In 5$s$,\,$p$ hybridized states with some contribution from Mn 3$d$ states. We note that similar hybridized states gives rise to a broad hump at 3.8 eV in the majority spin DOS. Therefore the satellite peak occurring in the XAS spectra can be primarily attributed to the Ni 3$d$ - In 5$s$,\,$p$ hybridized states. Similar satellite feature was observed earlier in Ni XAS of Ni$_2$MnGa~\cite{jakobs}. Based on theoretical calculations, it was assigned to a Ni 3$d$ - Ga 4$s$,\,$p$ hybridized peak in the unoccupied DOS~\cite{Entel}.

\begin{figure}[htb]
\centering \includegraphics[width=0.9\columnwidth]{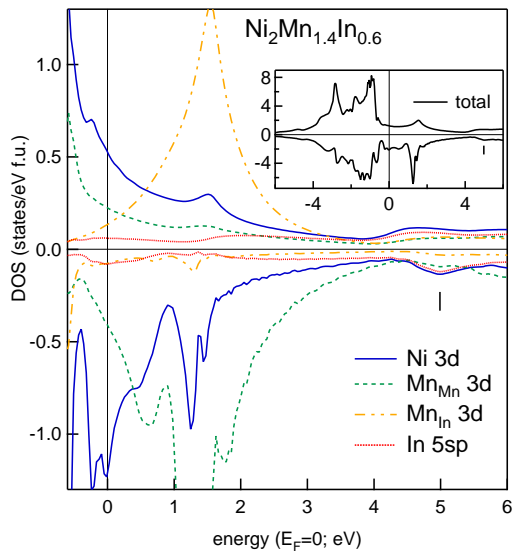} \caption{(Color online) Atom and spin projected DOS of Ni$_2$Mn$_{1.4}$In$_{0.6}$ in the tetragonal ($c/a$=~0.96) structure. Inset shows the total DOS over an extended region including the occupied DOS below $E_F$. The peak in the minority spin DOS at 1.25 eV is primarily due to Mn$_{\rm Mn}$ $3d$ states. In contrast, the peak at 1.55 eV arising from  Mn$_{\rm In}$ $3d$ states have majority spin character.} \label{dosIn15}
\end{figure}

Ni XAS for Ni$_2$Mn$_{1.4}$In$_{0.6}$ shown in Fig.~\ref{xas-Ni2MnIn}(b) also exhibits the satellite feature, albeit at higher energy. The satellite now occurs at 861.1 eV which is 6.5 eV above the L$_3$ edge. In fact a systematic shift in the satellite peak position is seen with change in temperature. At 310 K the satellite occurs at 859.8 eV and shifts to 861.1 eV at 15 K, following the transformation of Ni$_2$Mn$_{1.4}$In$_{0.6}$ from austenitic to martensitic phase\cite{supplement}. Such a shift was also observed in Cu doped Ni$_2$MnGa \cite{sroy}. Interestingly, EXAFS study of several Ni$_2$Mn$_{1+x}$In$_{1-x}$ compositions shows that the average Ni-Mn bond distance is shorter than Ni-In bond distance in the austenitic phase\cite{krp-apl} and this difference only increases upon martensitic transformation. Such local structural changes can result in increase in hybridization between Mn $3d$ and Ni $3d$ states and could be the reason for the shift in position of the satellite feature in Ni$_2$Mn$_{1.4}$In$_{0.6}$. This argument is further supported by photoelectron spectroscopy study of Mn rich Ni-Mn-Sn alloys that show the formation of Mn-Ni hybrid states near the E$_F$ upon martensitic transformation~\cite{kimura}. Our DOS calculations presented in Fig. \ref{dosIn15} show that Mn contribution to the total DOS in Ni$_2$Mn$_{1.4}$In$_{0.6}$ is more than that in Ni$_2$MnIn in agreement with experiment. This indicates an increased hybridization between Ni $3d$ and Mn $3d$ states in Ni$_2$Mn$_{1.4}$In$_{0.6}$.

Turning to the Mn XAS shown in Fig.~\ref{xas-Ni2MnIn}(c)\&(d), the overall multiplet features of both the compounds agree fairly well with many other Mn based Heusler alloys \cite{s1,s2,s3}. These features are considered to be a signature of localized $3d$ electrons \cite{s1,grab,thole}. Alternatively, a selective oxidation of Mn atoms can also result in multiplet structures \cite{s2,s3}. In either case, as has been shown in the supplementary text \cite{supplement}, occurrence of these multiplet do not affect our overall conclusion.


\begin{figure}
\centering
\includegraphics[width=\columnwidth]{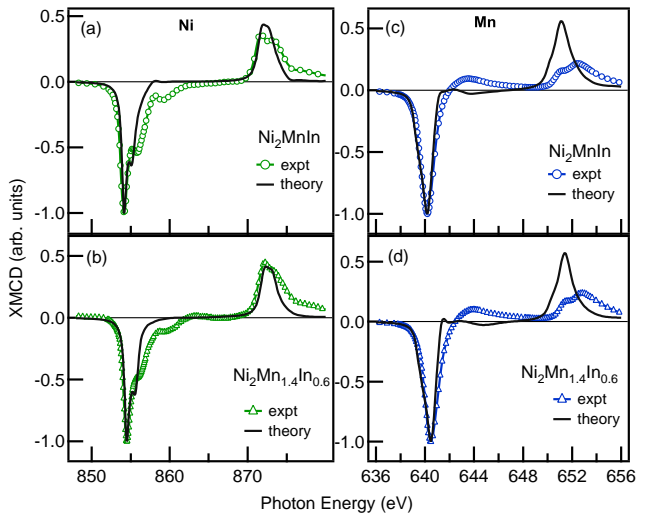}
  \caption{(Color online) Ni and Mn $L_{3,2}$-edge XMCD spectra of both the compositions measured at 15 K and compared with the calculated spectra. The peak heights of the experimental and calculated spectra are matched to unity for clear representation and clarity.}
  \label{xmcd}
\end{figure}

\begin{figure}
\centering
\includegraphics[width=\columnwidth]{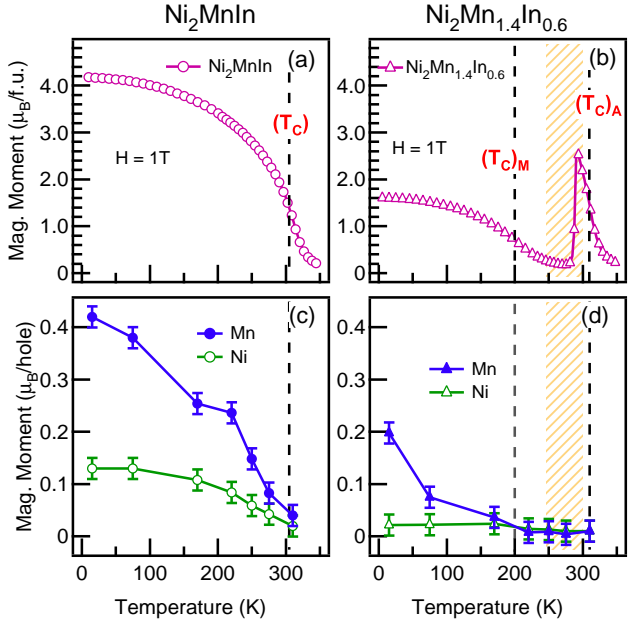}
  \caption{(Color online) Temperature variation of magnetization measured in the field of 1T for (a) Ni$_{2}$MnIn  and (b) Ni$_2$Mn$_{1.4}$In$_{0.6}$. Magnetic moments extracted from the present XMCD data for the two compositions is shown in (c) \& (d) respectively.The dashed vertical lines indicate magnetic ordering temperature and the region of martensitic transformation is depicted as cross-hatched area.} \label{moment}
 \end{figure}

The experimental and calculated XMCD spectra for Ni and Mn L edges of the two compositions are shown in Fig.~\ref{xmcd}. Ni$_2$MnIn  shows a robust dichroism signal at Mn-edge giving $\mu_{spin}\sim 3.7~\mu_B/atom$, while Ni gives $\mu_{spin}\sim 0.34~\mu_B/atom$ at 15 K.
The calculated values are in close agreement to experiment with $\mu_{spin}$= 3.46~$\mu_B/atom$ for Mn and 0.34~$\mu_B/atom$ for Ni. The total magnetic moment estimated from the present analysis is also in good agreement with magnetization measurements presented in Fig.~\ref{moment}(a). For Ni$_2$Mn$_{1.4}$In$_{0.6}$ the Mn and Ni magnetic moments estimated from XMCD are 1.45~$\mu_B/atom$ and 0.03~$\mu_B/atom$ respectively, giving a total moment of $\sim$1.5~$\mu_B$ which agrees well with magnetization measurement value of 1.6~$\mu_B$ as can be seen from Fig~\ref{moment}(b). From Fig.~\ref{moment}(c) it is seen that the  temperature dependence of the magnetic moment of Ni$_2$MnIn obtained from XMCD matches with the magnetization curve. However, in the case of Ni$_2$Mn$_{1.4}$In$_{0.6}$, presented in Fig. \ref{moment}(d), a striking observation that bring to fore the role played by Ni in building up its magnetic interactionsis that the spin moment of Mn and Ni is much smaller in comparison to that in Ni$_2$MnIn. While a lower moment of Mn can be reconciled as a sum of two antiparallel contributions arising from Mn$_{\rm Mn}$ and Mn$_{\rm In}$, it is not expected for Ni to have a lower magnetic moment unless, we consider the possibility that some of the Ni atoms, especially those that find themselves between Mn$_{\rm Mn}$ and Mn$_{\rm In}$ are aligned parallel to Mn$_{\rm In}$, whereas the Ni atoms between Mn$_{\rm Mn}$ and In align parallel to Mn$_{\rm Mn}$. Since Mn$_{\rm Mn}$ and Mn$_{\rm In}$ are antiparallel, the moments of the {\it in-between} Ni atoms are also antiparallel thus explaining almost zero moment of Ni. This is supported by the EXAFS study that indicated Mn$_{\rm In}$ atoms to be closer to Ni than In atoms~\cite{krp-apl}. Such shorter bond distance results in higher Mn-Ni exchange interaction \cite{sasi1}. The antiferromagnetic interaction between Mn$_{\rm Mn}$ - Mn$_{\rm In}$ that is mediated by Ni can thus be explained by the superexchange type indirect interaction proposed in Ref. \onlinecite{sasi}.

\begin{figure}
\centering
\includegraphics[width=\columnwidth]{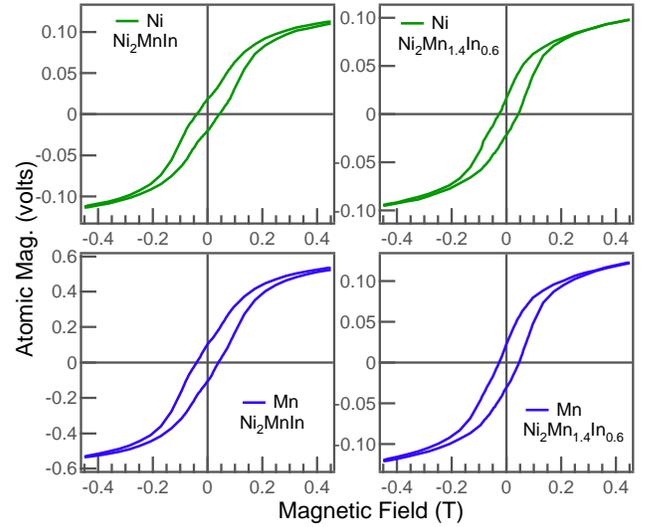}
  \caption{(Color online) Element specific hysteresis obtained from XMCD measurement (see text for details). Data are presented only at 15 K and in limited magnetic field range for clarity, although measurements were performed at several temperatures in the interval 310 K $\le$ T $\le$ 15 K and magnetic field +2T to -2T.}
  \label{hys}
\end{figure}

Finally, we present evidence for the participation of Ni in establishing AFM interactions in the martensitic phase of Ni$_2$Mn$_{1.4}$In$_{0.6}$. This crucial information is obtained from the measurement of element specific hysteresis loops, carried out within the XMCD setup. Here the incident energy was tuned to just above the L edge resonances of Mn and Ni and the sample current was monitored upon ramping the magnetic field (-2T to 2T). Thus the observed hysteresis loop reflects the magnetic contribution of the particular excited atom. The loops acquired at Mn and Ni L edges of Ni$_2$MnIn and Ni$_2$Mn$_{1.4}$In$_{0.6}$ at 15 K are shown in Fig. \ref{hys}. While the hysteresis loops obtained for Ni$_2$MnIn are symmetric about the origin of the graph, those obtained for Ni$_2$Mn$_{1.4}$In$_{0.6}$ are displaced to the right of the horizontal axis. Such asymmetry around the zero of magnetic field draws parallel with the exchange-bias effect observed in magnetization study \cite{pat,pab}. The assymmetry in the hysteresis loop arises due to presence of FM and AFM interactions even below (T$_C$)$_M$. However, the present case is a step ahead as it undoubtedly proves the participation of both Mn and Ni atoms in the AFM interactions taking place in the martensitic phase of Ni$_2$Mn$_{1.4}$In$_{0.6}$.

Ni$_2$MnIn, which crystallizes in L2$_1$ structure, has Ni atoms at the body centered position of CsCl type cubic sub-cell of which the corners are shared alternately by Mn and In. Therefore, there are Mn-Ni-In chains present along the [111] direction of the cube. In Ni$_2$Mn$_{1.4}$In$_{0.6}$, 40\% of In atoms are replaced by Mn leading to a formation of Mn-Ni-Mn chains along with Mn-Ni-In chains in the unit cell. Presence of local structural distortions in Ni$_2$Mn$_{1.4}$In$_{0.6}$ results in increased Ni $3d$ - Mn $3d$ hybridization. Evidence for an increase in such a hybridization can also be seen from our calculations discussed above.

\begin{figure}
\centering
\includegraphics[width=\columnwidth]{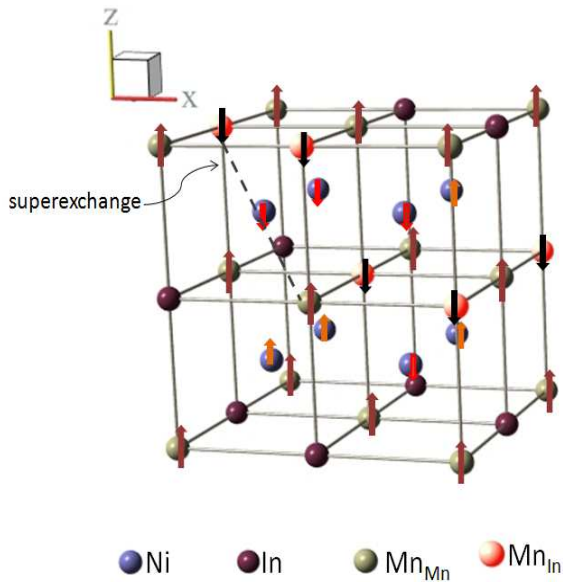}
  \caption{Schematic of atomic magnetic moments in Ni$_2$Mn$_{1.4}$In$_{0.6}$.}
  \label{SE}
\end{figure}

We propose that superexchange type interactions develops between Mn-Ni-Mn diagonal chains formed as a result of Mn occupying the In sub-lattice in addition to its own and the local structural distortion leads to the strengthening of the AFM interaction. A schematic of such an interaction is shown in Fig. \ref{SE}. The Ni atoms that find themselves in between Mn and In atoms have ferromagnetic moment, while those that are placed between two Mn atoms, align with their spins in the opposite direction. This reduces Ni moment drastically as every substituted In will affect the nearest neighbor Ni sites. This is also very clearly seen from the moment values extracted from XMCD measurements where the Ni moments are considerably reduced in Ni$_2$Mn$_{1.4}$In$_{0.6}$ as compared to that in Ni$_2$MnIn. If the antiferromagnetic interactions were purely RKKY type between Mn atoms, the Ni moment should not have decreased so drastically. The strong evidence of Ni participating in antiferromagnetic interactions is of~course the observation of shifted hysteresis loops in the XMCD measurements of Ni$_2$Mn$_{1.4}$In$_{0.6}$, which is akin to exchange bias effect as observed in magnetization measurements.

\section{Conclusions}
In summary, we have shown that the origin of AFM interactions present in the martensitic phase of Ni$_{2}$Mn$_{1.4}$In$_{0.6}$ lies in superexchange interactions between Mn atoms mediated by Ni. The XAS at Ni L$_{2,3}$ edges in Ni$_2$MnIn and Ni$_2$Mn$_{1.4}$In$_{0.6}$ indicates a substantial increase in hybridization between Ni and Mn atoms. This observation is further supported by spin polarized DOS calculated for the two compounds. As a result of increased hybridization, a redistribution of electrons taken place between the Ni~$3d$--Mn~$3d$, hinting that superexchange-like interactions are at play. Temperature dependent change in magnetic moments of Mn and Ni are also well mapped and emulates the magnetization curve obtained using magnetometer based measurements. The ultimate evidence for the participation of Ni in AFM coupling comes from the shifts seen in the hysteresis loop measurements carried out within the XMCD framework.

\section*{Acknowledgments} K.R.P and D.N.L acknowledge travel support from the Department of Science and Technology, Government of India through S. N. Bose National Centre for Basic Sciences, Kolkata. JASRI is thanked for beamtime at BL25SU, SPring-8, Japan (Proposal No. 2010A1040) and the help from Dr. T. Nakamura for these experiments is gratefully acknowledged. H. Ebert and M. Offenberger are thanked for useful discussions related to  the KKR calculations. CSIR, New Delhi is thanked for financial support under 03/EMR-II/1188 and for providing research fellowship to S.W.D.

\end{document}


\maketitle{\centerline {\em Supplementary material to the paper titled}
{\bf{\noindent Antiferromagnetic Exchange Interactions in Ni$_2$Mn$_{1.4}$In$_{0.6}$ferromagnetic Heusler alloy}}\\




\begin{figure}[h]
\epsfxsize=150mm \epsffile{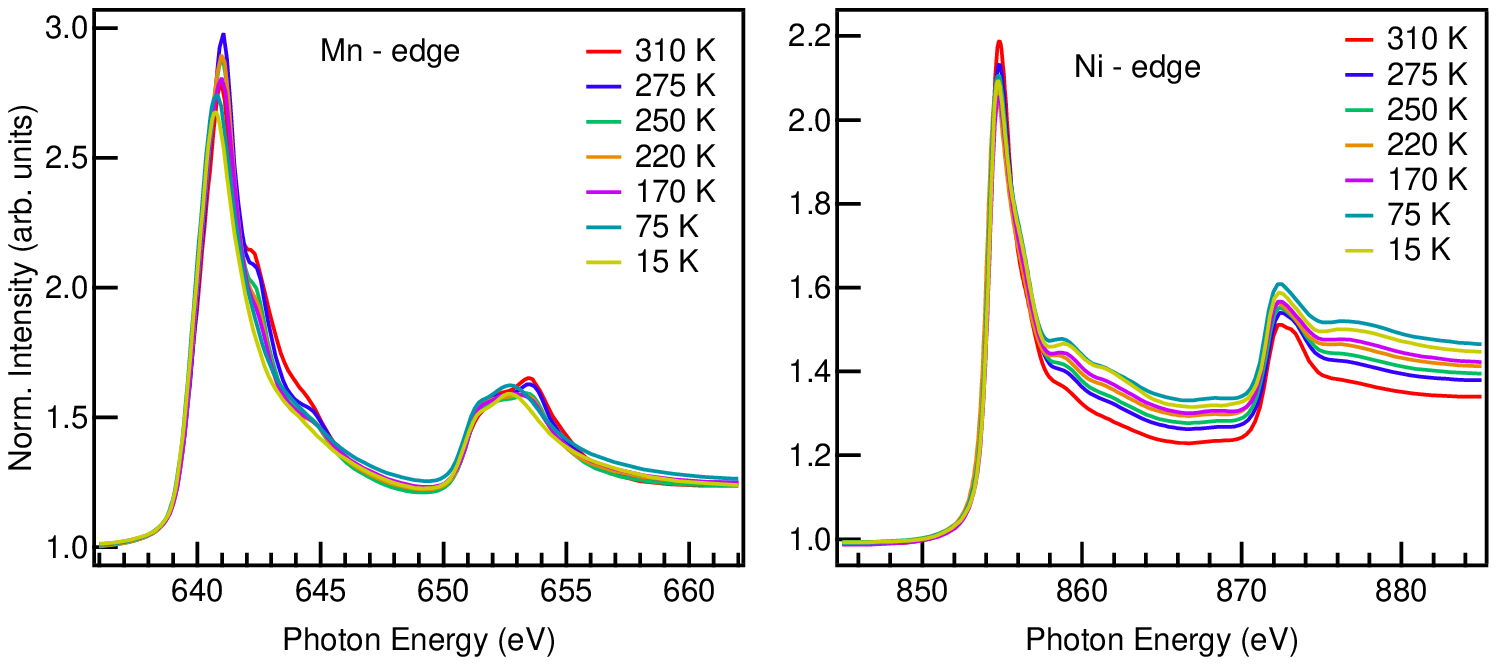}
\begin{center}
Figure S1: Temperature dependent XAS at Mn and Ni L edges in Ni$_2$MnIn
\end{center}
\end{figure}

XAS and XMCD spectra of Ni$_2$MnIn and Ni$_2$Mn$_{1.4}$In$_{0.6}$ were recorded in temperature interval 15K $\le T \le$ 310K at BL25SU beamline at SPring-8, Japan as per the details mentioned in the paper.

Figure S1 shows the temperature dependance of XAS plots at Mn and Ni L$_{2,3}$ edges in Ni$_2$MnIn. In order to normalize all the spectra, a common background was subtracted from the lower energy region, before the absorption jump and the spectra were normalized to a constant area under the curve. The doublet structure at the Mn L$_2$ and the multiplet structures seen in Mn L$_3$ edges of both the alloys are common features observed in many other Mn based Heusler alloys \cite{s1,s2,s3}. The multiplet features are considered to be a signature of localized
$3d$ electrons \cite{s1,grab,thole}. Alternatively, a selective oxidation of Mn atoms can also result in similar multiplet structures \cite{s2,s3}. Therefore it is difficult to assign one of the two causes for the origin of multiplet structures in Mn L edges reported here. However, even if these multiplets are due to selective oxidation of Mn, they do not affect the final conclusions as these do not contribute to the XMCD signal. In Fig. S2 the Mn spectra recorded for two different helicities of incident photons show hardly any difference between them at energy values where these multiplets are present and hence there is no contribution to the resulting XMCD spectra. Therefore the XMCD spectra results purely from ferromagnetic Mn atoms belonging to the Heusler alloy phase. On the other hand, the satellite feature observed on the higher energy side of Ni L$_3$ edge, at 859.2 eV has a special significance. As can be seen in Fig. S1(b), this feature is present at all the temperatures. This feature has also been reported in Ni$_2$MnGa and has been ascribed to Ni 3$d$-Ga 4$s,p$ hybridized states \cite{s4}. Likewise in the present context of Ni$_2$MnIn, the origin of the satellite lies in the hybridization of Ni 3$d$ and In 5$s,p$ states as also confirmed from our DOS calculations.

\begin{figure}[h]
\epsfxsize=100mm \epsffile{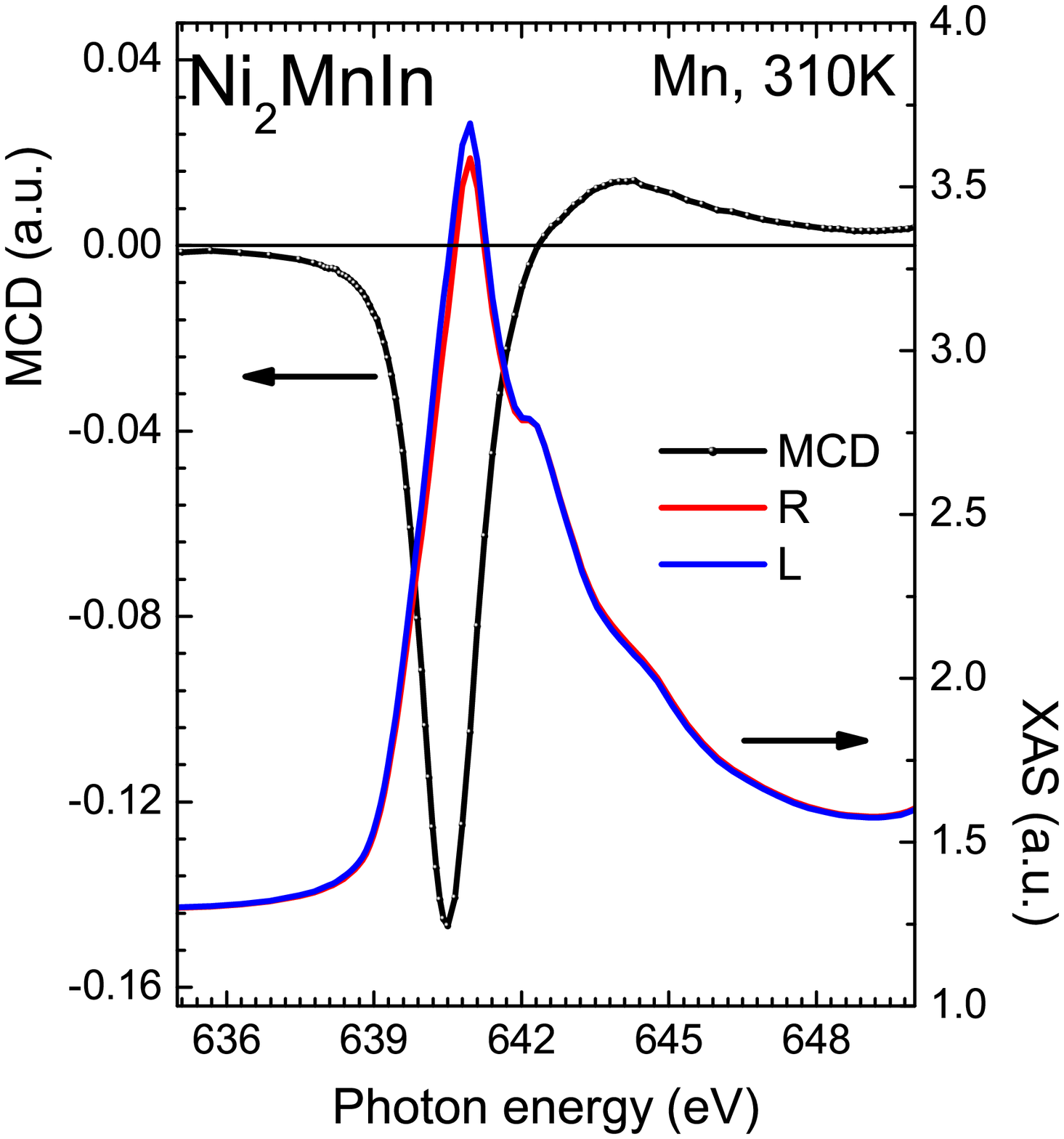}
\begin{center}
Figure S2: Mn L$_3$ edge spectra recorded for both right and left helicity photons and the corresponding XMCD
signal at 310K in Ni$_2$MnIn
\end{center}
\end{figure}

XAS data for Ni$_2$Mn$_{1.4}$In$_{0.6}$ recorded at various temperatures is presented in Fig. S3. The spectra for Mn and Ni L edges are quite similar to those in Ni$_2$MnIn. One noticeable change however, is in the position of the satellite peak above Ni L$_3$ edge. This satellite feature shifts to higher energy upon the alloy undergoing martensitic transformation. At 310K the satellite appears at 859.8 eV while in the martensitic phase, i. e. $T <$ 275K the peak shifts to 861.1 eV that is higher in energy by 1.3 eV.\\

\begin{figure}[h]
\epsfxsize=150mm \epsffile{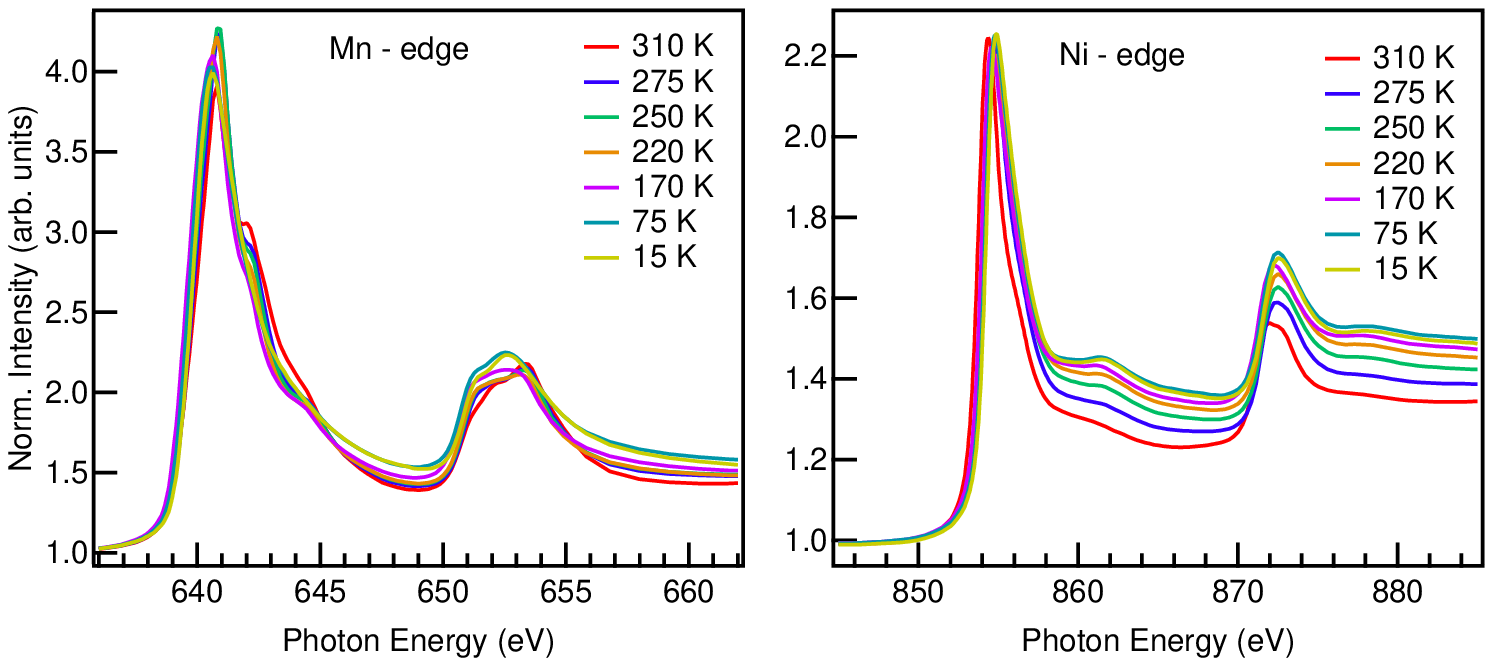}
\begin{center}
Figure S3: Mn and Ni L edges as a function of temperature in Ni$_2$Mn$_{1.4}$In$_{0.6}$
\end{center}
\end{figure}

\noindent Figure S4 exhibits temperature dependance of XMCD signal obtained at Ni and Mn L$_{2,3}$ edges in Ni$_2$MnIn. For both, Mn and Ni the XMCD signals increase steadily with decreasing temperature indicating build up of ferromagnetic moment. \\

\begin{figure}[h]
\epsfxsize=150mm \epsffile{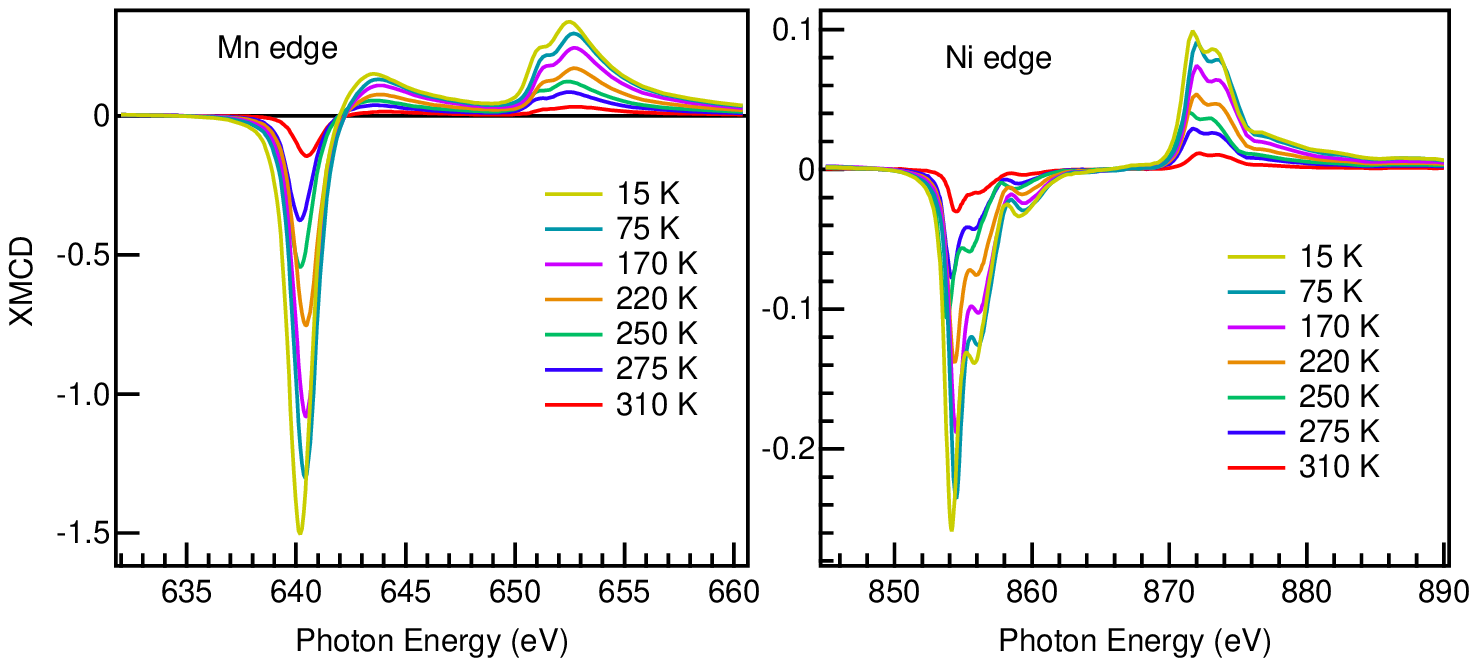}
\begin{center}
Figure S4: XMCD signals at Mn and Ni L edges at various $T$ in Ni$_2$MnIn
\end{center}
\end{figure}

\noindent Mn and Ni L$_{2,3}$ XMCD signal obtained in Ni$_2$Mn$_{1.4}$In$_{0.6}$ at various temperatures is shown in Figure S5. There is a noticeable drop in the XMCD signal of both, Mn and Ni as the temperature decreases from 310K to 275K indicating a reduction of magnetic moment of Mn and Ni upon martensitic transformation. Below 250K, the intensity of martensitic signal increases with decrease in temperature. This correlates well with the build up of magnetic moment seen in magnetization measurements.\\

\begin{figure}[h]
\epsfxsize=150mm \epsffile{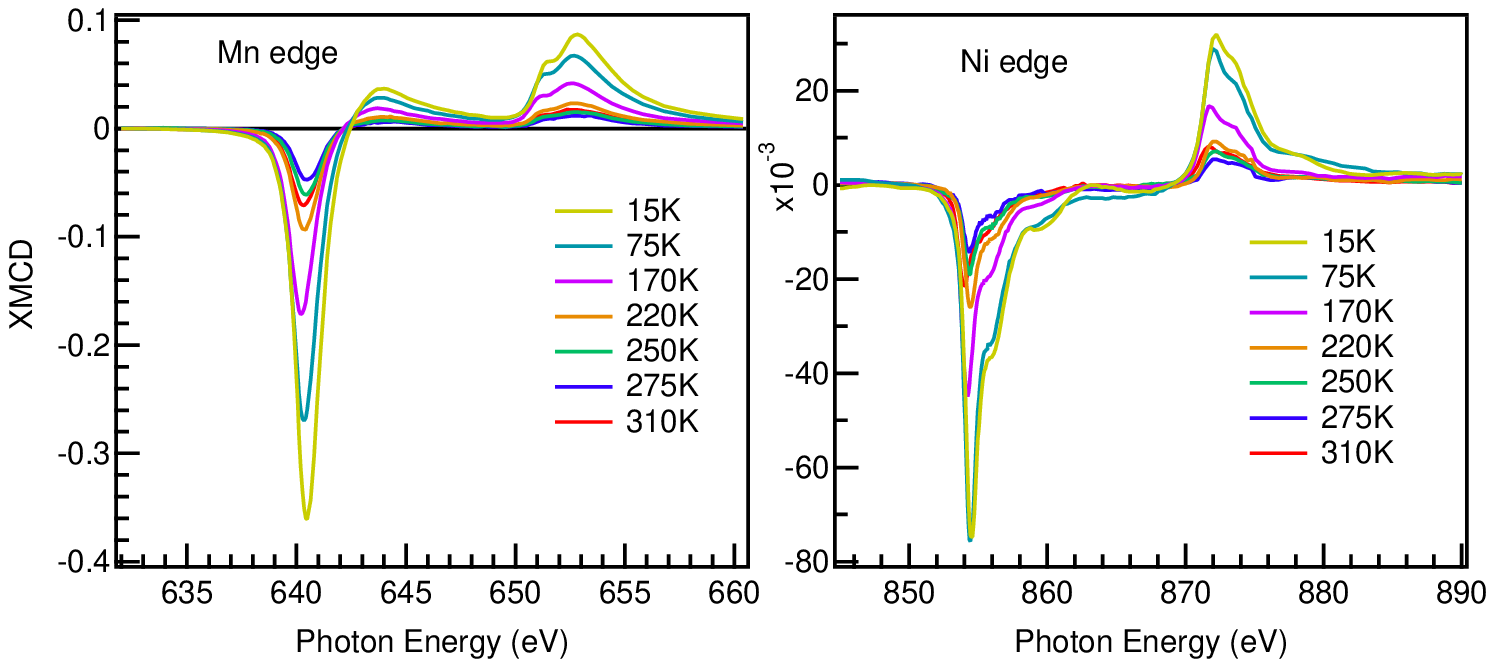}
\begin{center}
Figure S5: $T$ dependent XMCD signals at Mn and Ni L edges in Ni$_2$Mn$_{1.4}$In$_{0.6}$
\end{center}
\end{figure}